# Simple experimental procedures to distinguish photothermal from hot-carrier processes in plasmonics

Guillaume Baffou,[*,†] Ivan Bordacchini,[‡] Andrea Baldi[#,§] and Romain Quidant[‡,¶]

[†] Institut Fresnel, CNRS, Aix Marseille University, Centrale Marseille, Marseille, France

[‡] ICFO-Institut de Ciències Fotòniques, The Barcelona Institute of Science and Technology, 08860 Castelldefels, Barcelona, Spain

[#] DIFFER – Dutch Institute for Fundamental Energy Research, De Zaale 20, 5612 AJ Eindhoven, The Netherlands

[§] Vrije Universiteit Amsterdam, De Boelelaan 1081, 1081 HV Amsterdam, The Netherlands

[¶] ICREA-Institució Catalana de Recerca i Estudis Avanats, 08010 Barcelona, Spain

[*] e-mail: guillaume.baffou@fresnel.fr

**Abstract** - Light absorption and scattering of plasmonic metal nanoparticles can lead to non-equilibrium charge carriers, intense electromagnetic near-fields, and heat generation, with promising applications in a vast range of fields, from chemical and physical sensing, to nanomedicine, and photocatalysis for the sustainable production of fuels and chemicals. Disentangling the relative contribution of thermal and non-thermal contributions in plasmon driven processes is however difficult. Nanoscale temperature measurements are technically challenging and macroscale experiments are often characterized by collective heating effects, which tend to make the actual temperature increase unpredictable. This work is intended to help the reader experimentally detect and quantify photothermal effects in plasmon-driven chemical reactions, to discriminate their contribution from the one due to photochemical processes, and to cast a critical eye on the current literature. To this aim, we review, and in some cases propose, seven simple experimental procedures, which do not require the use of complex or expensive thermal microscopy techniques. These proposed procedures are adaptable to a wide range of experiments and fields of research where photothermal effects need to be assessed, such as plasmonic-assisted chemistry, heterogeneous catalysis, photovoltaics, biosensing and enhanced molecular spectroscopy.

# Introduction

Driving chemical reactions with plasmonic nanoparticles is a rapidly growing field of research, with potential applications of high economical and industrial impact. Localized surface plasmon resonances in metal nanoparticles can catalyze chemical reactions via optical near-field enhancement, heat generation, and hot charge carrier injection [1]. The latter mechanism, based on the use of non-equilibrium electrons and holes to activate redox reactions, was initially proposed in 2004, paving the way to a very active branch of research in plasmonics [2, 3]. Photon absorption in plasmonic metal nanoparticles results in the excitation of non-equilibrium electron-hole pairs with energy as high as a few eV. Such non-thermal, highly energetic charge carriers are coined *hot-carriers* in solid-state physics, because they markedly deviate from the thermalized Fermi-Dirac energy distribution of the free electrons in the metal. The transfer of these hot charge carriers from the nanoparticle to the surrounding molecular adsorbates or photocatalytic materials (such as $TiO_2$) is capable of driving electronic and chemical processes at the nanoparticle vicinity. Since 2010, there has been a sudden rise in the number of publications related to hot-carrier plasmonics, driven by seminal work from the groups of Moskovits [4, 5], Halas [6] and Linic [7, 8], among others, and envisioning applications in nanochemistry [8, 9], water-splitting [5, 7, 10], optoelectronics [6, 11] and photovoltaics [4, 12, 13, 14].

Depending on the application, different definitions of hot-electrons in plasmonics have been used, and some clarification has to be made before going further, to avoid confusion and ambiguity. After a photon is absorbed by a metal nanoparticle, a very energetic electron-hole pair is created, with an energy equal to the photon energy hν. This energy is shared between these two carriers with a ratio that depends on where the excited electron originates from within the conduction band [15]. These primary hot carriers are usually coined quasi-ballistic carriers [16, 17, 15]. Within a few tens of fs [18], the primary hot carriers thermalize with the other electrons of the metal through electron-electron inelastic scattering events. These subsequent thermalized charge carriers have a strongly reduced energy compared with the primary hot electrons, less than a few tenth of eV. However, they have also been called "hot" by a large part of the community, especially working with pulsed lasers, as such low energies still correspond to electronic temperatures on the order of a few 1000s of K. These thermalized, "warm" electrons [16] should be distinguished from the primary, quasi-ballistic hot electrons because of their lower energy and their longer lifetimes, of the order of picoseconds, dictated by multiple, sequential electron-phonon scattering processes. In hot-carrier assisted plasmonic chemistry, only primary hot electrons have enough energy to contribute to chemical reactions.

The actual involvement of hot-carriers in several chemistry experiments has been recently questioned, with the proposition of alternative mechanisms, such as direct photoexcitation of hybrid particle-adsorbate complexes [19, 20, 21], or simple heat generation [22, 23, 24]. Indeed, the further thermalization of these excited carriers via electron-phonon scattering leads to heating of the entire nanoparticle and further heat diffusion to the surrounding reaction medium, suggesting that photothermal effects may also contribute to the observed reactivity enhancement [25].

The main concern with primary hot-carriers resides in their very short lifetime. They thermalize via electron-electron scattering within a time scale $\tau_{\text{e-e}}$ of a few tens of fs for gold [18], making any interaction with the surrounding environment a low probability event. The time-average number of primary hot electrons generated in a single nanoparticle under illumination can be quantified using this simple expression:

$$< N_{\text{hot e-}} > = \frac{\sigma_{\text{abs}} I \tau_{\text{e-e}}}{h\nu} \qquad (1)$$

where $\sigma_{\text{abs}}$ is the absorption cross-section of the nanoparticle, I the irradiance (power per unit area) of light, $\tau_{\text{e-e}} \sim 50$ fs and $h\nu$ the photon energy. For a gold nanosphere 50 nm in radius ($\sigma_{\text{abs}} = 2 \times 10^4$ nm$^2$) illuminated at 530 nm with $I = 5 \times 10^4$ W/m$^2$ (a typical value from the literature [7, 8, 9, 26, 27, 28]), the time-average number of hot electrons in the nanoparticle under steady state illumination $< N_{\text{hot e-}} >$ is around $10^{-4}$. This low number means that for irradiances typically used in plasmon-assisted photochemical experiments, primary hot charge carriers are only available for ~0.01% of the time, i.e., on very brief occasions. Under CW illumination, a hot carrier always thermalizes before the absorption of the next photon, such that no primary hot carrier population exist, a picture that only applies under pulsed illumination where 1000s of photons can be absorbed during $\tau_{\text{e-e}}$. However, under the conditions mentioned above, the nanoparticle still absorbs around 3 billion photons per second, generating 3 billion primary hot electron-hole pairs. Thus, the very small lifetime of the primary hot carriers is unfavorable but does not necessarily mean the impossibility of noticeable hot-carrier assisted processes, a priori. The community is aware of this issue and several recent studies directly analyze the respective contributions of hot electrons and photothermal effects, as shown in

Figure 1. Nevertheless, although plasmonic nanoparticles are excellent light to heat converters, the associated temperature increase is often difficult to predict and to measure.

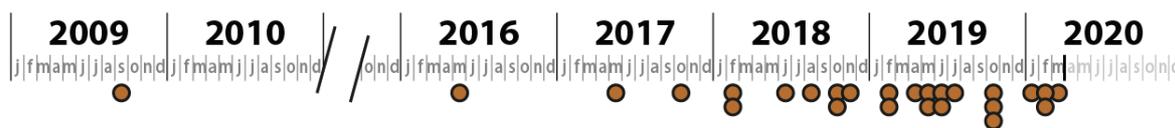

*Figure 1 : Publication timeline of articles focusing on photothermal effects in hot-carrier assisted plasmonics.* 2009: [29], 2016: [30], 2017: [26, 31], 2018: [32, 33, 34, 35, 23, 36, 24], 2019: [22, 37, 27, 38, 39, 40, 41, 42, 23, 43] [44], 2020: [45, 46, 47, 48].

This article is intended to help experimentalists discriminate between thermal and non-thermal effects in plasmon-driven chemical processes. To this aim, we propose seven simple experimental procedures that avoid the use of complex or expensive thermal microscopy techniques, which may sometimes be inaccurate [49, 50]. These procedures are described hereinafter and critically illustrated with some practical examples from the literature. In the last section, we also provide some practical guidance on how to avoid common pitfalls when using numerical simulations to estimate photothermal effects in plasmonic systems, highlighting the importance of collective photothermal effects in plasmonics.

## Procedure # 1: Varying the illumination power

In the case of a photochemical process in plasmonics, such as near-field enhancement of photochemical reactions or hot charge carrier assisted redox reactions at the nanoparticle surface, the rate $\eta$ [mol.s$^{-1}$] of chemical transformation is proportional to the rate of incident photons and therefore to the incident light power impinging on the sample. This assumption holds true for CW illumination under moderate light power and may deviate toward a superlinear dependence for very high power [51] or under fs-pulsed laser illumination due to multiphoton absorption [52, 53, 54, 55].

The case of a photothermal process is however different and should not feature such a linear dependence. Within a good approximation, the temperature increase of a system due to light absorption is also proportional to the optical power impinging onto the sample. However, the rate constant $K$ of a chemical reaction typically follows an Arrhenius-type temperature dependence, $K = A \exp(-E_a/RT)$, where $R$ is the gas constant, $T$ is the temperature, $E_a$ is the molar activation energy and $A$ the pre-exponential constant factor. Consequently, in the case of a photothermal process, the rate of chemical transformation follows an exponential dependence on the illumination power.

These two different dependences of the reaction rate on the incident optical power offer a simple means to discriminate a photothermal effect from a photochemical effect by plotting the measured chemical reaction rate (or any readout of the amount of reaction products) as a function of the light source power. A linear dependence would indicate a photochemical process while an exponential increase would rather be the signature of a photothermal effect.

Despite such a simple and intuitive reasoning, caution must be taken with this first trick. For example, in 2011, in a seminal article reporting on the plasmon-assisted epoxidation of ethylene [8], the authors proposed a hot-electron mechanism to explain their experimental data. A photothermal mechanism was discarded by kinetic isotope effect measurements and by using Procedure #1: a linear (non-exponential) dependence was found between the laser power and the rate of reaction, as indicated by the linear fit in Figure 2a. Interestingly enough, the same experimental results can be interpreted assuming a purely thermal mechanism as shown in Figure 2b, where the original data is fitted with the Arrhenius equation given above, assuming a linear dependence of the sample temperature on the laser power.[1] The fitted parameters indicate that, assuming a purely photothermal process, the laser power range used in the experiment corresponds to a sample temperature increase $\delta T \sim 20$ K at $T_0 = 450$ K, i.e., a temperature variation of only 4%. For such a narrow temperature window, the Arrhenius law is nicely approximated by a linear dependence. The same issue can be observed in a very recent study of the production of ammonia in the presence of Au-Ru nanoparticles [56], as reproduced in Figure 2c-d.[2]

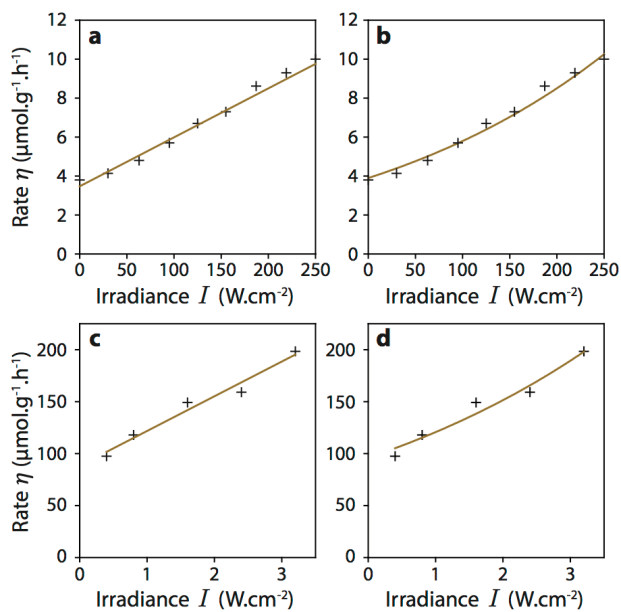

Figure 2: (a) Chemical rate of ethylene epoxidation as a function of light irradiance fitted using a linear law, as reported in Ref. [8]. (b) Same experimental data fitted using the Arrhenius law, as pointed out in Ref. [31], also showing good agreement. (c) Ammonia production rate as a function of light irradiance, fitted using a linear law as reported in Ref. [56]. (d) Same experimental data fitted using the Arrhenius law, also showing good agreement.

The above analysis shows that varying the laser power to deduce the relative contributions of photothermal and photochemical effects in plasmon-driven processes can only be done if the laser power is varied across a statistically significant range, typically leading to variations of the reaction rate over several orders of magnitude, not just by a factor of 2 or 3. Ideally, an experimental dataset should be large enough to be fitted with a superposition of an exponential (photothermal) and a linear (photochemical) terms. In practice, however, this is often difficult, as such a large range of chemical rates would entail the use of either extremely sensitive measurements of reaction rates at low powers or very high laser powers. Furthermore, it is possible that at high optical irradiances, the mechanism of the chemical process under study might change, due to the activation of alternative reaction pathways at high temperatures, or to non-linear optical effects. Under high intensity illumination and high temperature increases, convection effects in the gaseous or liquid surrounding medium could also occur. As convection favors heat removal, it could result in a sub-linear increase of the temperature and a non-exponential increase of the reaction rate with illumination intensity, despite a purely photothermal mechanism. Also, varying the rate of chemical reactions over multiple orders of magnitude may lead to additional complications due to changes in the catalyst surface coverage and therefore its activity and selectivity, or to saturation effects due to mass transport limitations. Note that a mass transport limitation would yield a damping of the chemical rate as a function of the illumination power, not an exponential increase, so it cannot be confused with a photothermal effect. For all of the above reasons, procedure #1 can only be used to discriminate photothermal from photochemical effects for well-characterized catalytic reactions over stable metal nanoparticles and preferably in conjunction with additional independent methods [51].

---

[1] The fitting function is $K = A \exp(-E_a/R(T_0 + cI))$, where $E_a = 85$ kJ/mol, $T_0 = 450$ K, $A = 2.9 \times 10^{10}$, $c = 80$ K.cm²/W.
[2] Same fitting function, where $E_a = 81.1$ kJ/mol, $T_0 = 293$ K, $A = 2.80 \times 10^{16}$, $c = 2.04$ K.cm²/W.

In 2012 a super-linear dependence (rate $\propto$ power$^n$, with n > 1) above a certain laser threshold was reported [51]. This observation was explained as a further confirmation of hot electron contribution through a DIMET (desorption induced by multiple electronic transitions) process on metal [57]. However, significant DIMET normally requires the use of a femtosecond-pulsed laser illumination [52, 53, 54, 55], which contrasts with the CW illumination used in the experiment. Moreover, these apparent super-linear increases of the reaction rate could also be fitted with a single exponential (Arrhenius) law with a convincing agreement [22], suggesting a photothermal mechanism.

Figure 2 offers the opportunity to discuss another relevant point. These studies reveal that these photocatalytic reactions also occur in the absence of light ($\eta \neq 0$ for $I = 0$, see also Figure 4). A residual catalytic activity indicates that the exponential factor of the Arrhenius law is already favorable in the dark at the operating temperature $T_0$, which naturally makes the system already exponentially sensitive to temperature. In order to more easily rule out photothermal effects, it would be interesting to study processes characterized by $\eta = 0$ in the absence of light and $\eta \neq 0$ under illumination. Even if such reactions may not be industrially relevant, this approach would be relevant from a fundamental perspective.

## Procedure # 2: Varying the light beam diameter

Instead of varying the illumination power, varying the light beam diameter can also provide valuable information. As shown further on, this procedure only applies for reactions occurring at the surface of a solid catalyst, such as a substrate covered with nanoparticles [6, 40, 58, 59] or an optically thick pellet [36]. It does not apply for photochemical reactions occurring on nanoparticles suspended in a liquid [35], where heat diffusion is more complex.

There are two common approaches to varying a light beam diameter on a sample plane: the constant-irradiance (power per unit area) approach and the constant-power approach. In the first case, the beam size is adjusted using a diaphragm on the beam path (Figure 3a) and the number of photons is thus proportional to the area of the sample under illumination. In the second case the beam is defocused to vary the beam size on the sample (Figure 3b) and the number of photons impinging on the sample is therefore constant.

Let us first consider how these two modes of illumination affect a light-induced process when it is photochemically driven. For a photochemical process, the reaction rate is proportional to the rate of incident photons, as mentioned in the previous Procedure #1. In the constant-irradiance mode, the reaction rate is thus supposed to be proportional to the area of the light beam impinging on the sample surface, while in the constant-power mode, no beam-size dependence is expected since the rate of photons impinging onto the sample is constant. Thus, depending on how the illumination beam diameter is varied, the photochemical rate features radically different variations. Note that this reasoning makes no assumption on the sample thickness. For this reason it applies not only for particles deposited on a flat substrate [6, 40, 58, 59], but also for thick samples, e.g. made of compacted powders or pellets [36].

Things are markedly different if the reaction is photothermally-driven. With the constant-irradiance approach, the temperature increase is proportional to the beam diameter: $\delta T \propto R_{\text{beam}}$ [60]. Thus, when opening the diaphragm, the reaction rate increases not only due to the enlarged irradiated area (like for a photochemical process), but also due to a higher temperature increase. With the constant-power approach the temperature increase is inversely proportional to the beam radius: $\delta T \propto 1/R_{beam}$ [60]. The smaller the beam, the higher the temperature. Thus, the rate of chemical reaction has no reason to be independent of the beam radius anymore, like with a photochemical process. In both cases (constant power and constant irradiance), the dependence of the chemical rate on the beam radius results from a subtle interplay between the variations of temperature and illuminated area. If one assumes moderate temperature variations leading to an Arrhenius law that resembles a linear law (like in Figure 2 [8]), one gets the dependencies summarized in Table 1. For both modes of operation, these relationships systematically differ for photochemical and photothermal processes. Investigating these dependences by varying the illumination diameter therefore appears as an efficient means to elucidate the underlying mechanism, or at least to show that the underlying process is not purely photochemical. These dependences of $\delta T$ on the beam radius assume that the heat produced by light absorption in the catalyst is efficiently dissipated via an infinite surrounding medium (just like Equation 3, further on), as is typically the case with a solid photocatalytic substrate. These dependencies also assume a two-dimensional heat source [60], and therefore a two-dimensional light-absorbing medium. If the absorbing medium is 3D or optically thick, like with a pellet, these dependences are still valid, provided that the heat source remains effectively 2D. This happens when the light penetration depth into the sample is small compared to the beam size. In practice, this condition is generally valid with an optically thick substrate, such as pellets, since they are highly scattering and absorbing by nature.

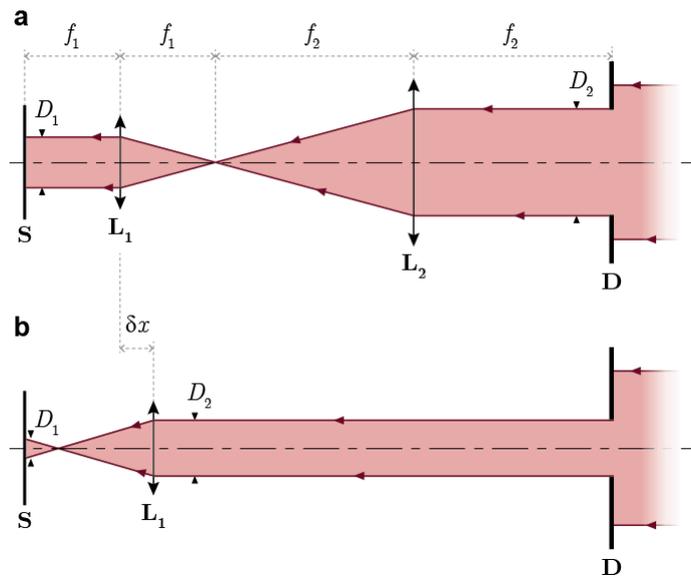

Figure 3: (a) Constant irradiance setup: 4f optical configuration enabling the setting of an illumination diameter $D_1$ at the sample location **S**, by adjusting the diaphragm diameter $D_2$, according to the relation $D_1 = D_2 \cdot f_1/f_2$. Note that for the 4f configuration to be properly used, the optical power density impinging on the diaphragm **D** has to be uniform, not Gaussian. (b) Constant power setup: optical configuration, similar to (a) when removing lens **L₂**, enabling the setting of an illumination diameter $D_1$ at the sample location **S**, by adjusting the displacement $\delta x$ of **L₁**, according to the relation $D_1 = \delta x \cdot D_2/f_1$. Note that **L₁** can be the objective lens of a microscope.

Table 1: Dependence of the reaction rate variation $\delta \eta$ on the light beam radius $R$.

|  | process | |
| --- | --- | --- |
| mode | photochemical | photothermal |
| constant power | $\delta \eta$ constant | $\delta \eta \propto R$ |
| constant irradiance | $\delta \eta \propto R^2$ | $\delta \eta \propto R^3$ |

We noted above that this procedure could be applied for solid samples, typically catalysts and nanoparticles dispersed on a planar substrate [6, 40, 58, 59] or pellets [36], in contact with a gas or a liquid phase, for which a heat source would remain two-dimensional. Indeed, the case of nanoparticles and reactants suspended in solution is more complex and cannot be faithfully investigated using this procedure. Similar dependences on the radius of an illuminated sphere could be derived for a three-dimensional heat source, ($\delta T \propto 1/R_{\text{beam}}$ for the constant power approach and $\delta T \propto R_{\text{beam}}^2$ for a constant light power density approach[3]), but these dependences still assume a heat diffusion that efficiently occurs over an infinite surrounding medium, i.e., no accumulation of heat within a thermally insulated vessel. This assumption is rarely valid for plasmon driven catalytic reactions in solution [61, 62], where the surrounding medium (sample holder, air around) may have a smaller thermal conductivity than the reaction medium itself (the liquid). This would lead to heat accumulation and uniformization within the whole liquid by heat conduction and convection, making any temperature estimation significantly more difficult than with a solid catalytic substrate, usually involving efficient surrounding conductive medium, such as a stainless steel chamber.

Interestingly, such a trick was previously mentioned by the group of Moskovits in 1994 in the context of photoemission measurements [63], although it was, in that case, rather used to discriminate between one-photon and two-photon processes. To our knowledge, this procedure has not yet been used to discriminate photothermal from photochemical effects in plasmon-assisted chemical reactions. For instance, it could have been relevant to studies such as Ref. [9], by measuring the rate enhancement of $H_2$ dissociation on gold nanoparticles as a function of the illumination area.

## Procedure # 3: infrared (IR) and thermocouple measurements

In practice, despite the sub-wavelength nature of the heat sources in plasmonics, *nanoscale* temperature measurements are not always required to properly estimate the temperature increase in a plasmonic reactor. In most experimental conditions, where the illumination spot size is much larger than the average interparticle distance, the illumination of a large number of particles at the same time gives rise to collective thermal effects, effectively suppressing nanoscale temperature inhomogeneities and leading to macroscopically homogeneous temperature distributions [60, 64, 35] (see last section of this article). Under these conditions, if the reactive surface is accessible to be imaged with an IR camera, infrared thermal measurements are certainly an excellent approach to monitor temperature variations of the sample.
However, as black body radiation depends not only on the temperature but also on the emissivity of the material, a proper calibration of the reactive medium is critical for reliable measurements. The non-reliable determination of the sample's emissivity has already been put forward as a possible source of errors in hot-carrier assisted plasmonic chemistry [36, 38, 39, 42]. In any case, one should avoid relying solely on a theoretical estimate of the emissivity based on the nature of the mixed materials covering the surface. Experimental measurements have to be performed. However, care has to be taken in particular with plasmonic samples, as metals are IR reflective and have thus very low emissivities, typically of the order of ~0.1, making IR temperature measurements even less reliable. Also, seeking a given emissivity not only assumes its spatial, but also its spectral uniformity, which is not always the case for photonic substrates [65]. A well-known procedure within the heat transfer community consists in determining an effective (spectrally averaged) emissivity experimentally, by uniformly heating the sample at different, well-defined temperatures, for instance within an oven or on a hot plate [66]. This procedure could be useful in case physical thermal contact of the sample with a thermocouple is problematic under operating conditions [67] (see discussion below).

---

[3] In this estimation, the 3D heat source consists of a sphere of radius $R_{\text{beam}}$, surrounded by an infinite, conductive surrounding medium.

Infrared cameras have been used, for instance, in experiments on plasmonic-assisted nanochemistry in gas phase [68] and in the study of plasmon-driven nanoparticle syntheses [35], revealing a significant temperature increase due to collective photothermal effects. A recent work reported on the use of an IR camera to monitor the temperature in heterogeneous catalysis in gas phase on Ru-Cu nanoparticles, where temperature increases larger than 100°C have been measured under normal illumination conditions [36]. The authors used a KBr window, transparent to the infrared, which is a requisite for reliable temperature measurements. It is important to underline that using an infrared camera is not suited for experiments where the reactive area is immersed in a liquid. In this case, the IR camera would probe the temperature of the surface of the liquid rather than the one at the reactive sites. More generally, the medium between the reactive area and the IR camera should not absorb IR light.

Alternatively, thermal measurements on macroscopic systems can be performed using thermocouples, which are ideally suited to measure the temperature in macroscopic three-dimensional samples [69]. This approach, however, has to be used with some caution as several problems can lead to an incorrect estimation of photothermal effects. First, one has to make sure that light does not directly impinge on the thermocouple, to avoid heating it directly. Second, physically contacting the substrate to a thermocouple can affect the local heat dissipation, which can in turn prevent reliable temperature measurements [67]. Finally, the thermocouple has to be put as close as possible to the reactive medium and in good thermal contact with it otherwise the temperature increase may be largely underestimated. This has been the case in some reported works on heterogeneous catalysis [9, 28] that used a commercial device (Harrick HVC-MRA-5) that was not meant to be heated with a laser (the window is rather meant to perform Raman measurements *in situ*). In this commercial system, conceived to be uniformly heated by an electrical current, the built-in thermocouple is not positioned within the device but away from the reactive area. The problem here is that light-driven plasmonic heating generates a non-uniform temperature increase within the device, localized on the reactive area. The temperature increase under illumination is thus underestimated by the remote thermocouple location. This problem was later pointed out by the authors themselves [36] as well as by other groups [24, 32] using the same commercial thermal reactor.

So far we have only focused on discriminating photothermal and photochemical processes in plasmon-driven chemical reactions. Another application that can benefit from a similar use of direct temperature measurements with thermocouples or IR cameras is plasmon-assisted photovoltaics. The presence of gold nanoparticles integrated into pn or Schottky junction solar cells has been shown to increase their photocurrent [70, 71, 72, 73, 6]. Three enhancement mechanisms have been proposed [74, 75]: more efficient light-trapping in the semiconductor, optical near-field enhancement and a hot-carrier injection from the plasmonic nanoparticles to the junction. Interestingly, the short-circuit current $I_{SC}$ of pn and Schottky junctions is increasing with temperature [76, 77]. Thus, an increase of the measured current in a solar cell could *a priori* originate from a photothermal effect. The temperature induced variation of $I_{SC}$ is in general extremely small, but extremely small current variations have been reported in plasmonic solar cells, so small (a few nA) that they had to be measured with a lock-in detection [6]. Thus, it is important to measure the temperature in such experiments to clearly identify the origin of the current increase. As the illuminated area of a solar cell is neither immersed in a liquid, nor sealed in a chamber, IR measurements represent a good option in this field of research, provided the emissivity of the solar cell is determined. More importantly, in such studies, one should not simply measure the short-circuit current, as often observed, but also the open circuit bias, or even the full $I - V$ characteristics like in Ref. [72] (see Figure 4a). Indeed, while a pure photothermal effect would lead to an increase of the short-circuit current, it would also cause a larger decrease of the open-circuit bias and a reduction of the cell filling factor, as sketched in Figure 4b. To our knowledge, thermal effects have never been considered as a possible mechanism driving photocurrent enhancements in plasmonic solar cells.

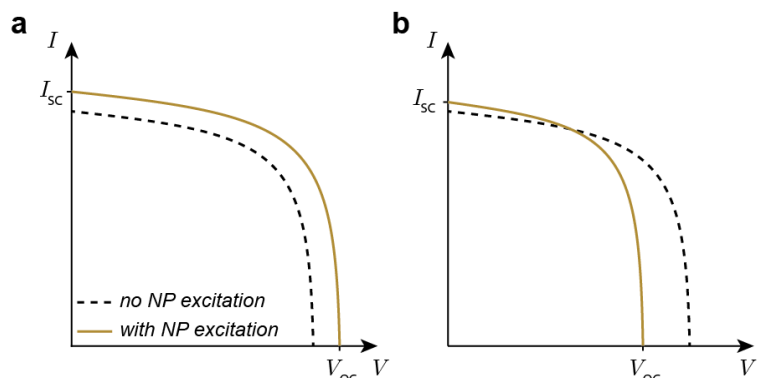

Figure 4: (a) Theoretical change of a solar cell $I - V$ characteristics expected if plasmonic nanoparticle excitation is actually improving the efficiency. (b) Effect of a temperature increase on a solar cell characteristics (pn or Schottky junction), also associated with an increase of the short circuit current $I_{SC}$ but with a decrease of the filling factor, and thus of the efficiency. These plots highlight that, to claim a positive effect of plasmonic nanoparticles on a solar cell efficiency, the full $I - V$ curve should be measured rather than only $I_{SC}$.

## Procedure # 4: Minding time scales

The time scale of a thermal process can be estimated using the simple expression $\tau = L^2/D$, where $L$ is the characteristic length scale of the heated area and $D$ is the thermal diffusivity of the surrounding medium, through which heat escapes. The parameter $\tau$ is the characteristic time to reach the new steady state temperature distribution following a heating perturbation. The thermal diffusivity is often difficult to estimate because the surroundings are usually not uniform (sample holder, catalyst support, flowing reactants and products, reaction vessel, etc.), but an order of magnitude can sometimes be obtained by considering average thermal diffusivities. In most cases, when illuminating a macroscopic sample (say one-inch in size), the kinetics of the temperature increase until reaching the steady state can be on the order of a few seconds to a few minutes. This fundamental difference in time scales can in principle be used to effectively discriminate between photothermal and photochemical effects: an instantaneous increase of the chemical rate indicates a pure photochemical effects [35], while a slow increase would rather suggest a photothermal effect, although caution must be used with the latter case as explained hereinafter. Note that similar information can be obtained by performing the opposite experiment of turning off the illumination and measuring the reaction rate decay with time.

Two examples taken from the literature on ethylene epoxidation and hydrogen dissociation are presented in Figure 5. In both cases, the reaction rate increases over one to several minutes. Thus, these measurements cannot be considered as instantaneous and used to ascertain a photochemical effect. However, albeit consistent with a photothermal effect, it does not mean that the underlying processes are necessarily photothermally-driven. Indeed, the measurement technique of the reaction rate may be endowed with some delay, for instance due to the diffusion of the products to the chemical sensor. For this reason, for this kind of experiment, a precise estimation of the response time of the chemical sensor should be determined, using for instance an inert molecular species as a tracer.

Only with this setup calibration could this procedure be applied to evidence or not the occurrence of photothermal effects.

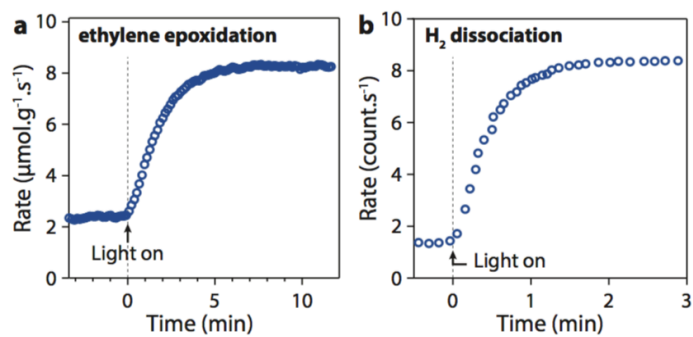

Figure 5: Time evolution of the increase in catalytic rate after switching on the illumination for two different plasmon-enhanced chemical reactions: (a) ethylene epoxidation on silver nanoparticles (data reproduced from Ref. [8]) and (b) hydrogen dissociation on gold nanoparticles (data reproduced from Ref. [9]).

## Procedure # 5: Calibrating with bubble formation

Upon optically heating plasmonic nanoparticles in a liquid environment, there necessarily exists a light power threshold where liquid-gas phase transition occurs, leading to the formation of one or several gas bubbles that can be easily visualized, for instance by optical microscopy [78, 79]. The formation of gas bubbles is therefore a direct indication of the existence of photothermal effects. By comparing the power required to generate a bubble with the one used in the typical experimental conditions, and assuming a linear temperature-power dependence [78], one can therefore estimate the presence and magnitude of photothermal effects. For instance, if a bubble appears upon only increasing the illumination power by a factor of two compared with the normal illumination conditions, then significant heating must be occurring. Conversely, if increasing the power by, e.g., a factor of 100 does not generate bubble formation, photothermal heating is most likely negligible. When ramping up the illumination power from the experimental conditions to those where bubble are being formed, the illumination spot size should remain constant. As we will discuss more extensively in the last section of this article, photothermal effects are extremely sensitive to how many particles are illuminated at the same time.

One should also be careful when using this procedure to estimate the sample temperature, since the temperature threshold is not necessarily the boiling point of the surrounding liquid. When working on glass substrate and by heating a confined volume through the objective lens of a microscope, the temperature threshold for bubble formation can easily and even systematically reach approximately 200°C in water [78]. This uncertainty on the liquid boiling point may add an error to the temperature estimation of up to a factor of 2. However, even such a high uncertainty could not be problematic if the aim is to qualitatively discard or confirm the occurrence of photothermal effects. Note that other phase transitions, such as the metal-insulator transition in vanadium dioxide or the condensation in thermotactic polymers, could in principle be used to calibrate photothermal effects in a photochemical setup.

## Procedure # 6: Comparing the effects of two polarizations

Some asymmetric metal nanostructures feature an optical near-field enhancement distribution that is highly dependent on the incident light polarization, while the absorbed power, and therefore the temperature increase, does not (see Figure 6). This kind of nanostructure was introduced in 2017 and coined photothermal isosbestic nanostructure (PIN) [80]. A sample made of PINs could thus be used, in general, to distinguish between an optically driven process and a photothermally driven process in chemistry. It would suffice to measure the chemical reaction rate as a function of the incident light polarization. No variation would indicate a photothermal process, while variations following the near-field enhancement factor would rather evidence a photochemical process.

This technique is not suited for nanoparticles randomly deposited on a substrate as usually observed in plasmon-enabled chemistry experiments, since all the plasmonic structures should be aligned along the same direction. It is however possible to use this procedure with samples made of plasmonic nanostructures fabricated by nanolithography techniques, such as e-beam lithography, substrate conformal imprint lithography [81], nanosphere or colloidal hole-mask lithography [82] or shrinking-hole colloidal lithography [83]. Even for nanolithography techniques that do not allow the fabrication of nanoparticles over macroscale areas, modern ultrasensitive gas chromatography / mass spectrometry (GC-MS) techniques can detect catalytic products of as few as 10 plasmonic particles [84, 85].

In this approach, the absorption cross sections of the nanoparticles predicted by numerical simulations might not exactly correspond to experimental observations. In particular, imperfection of the lithographic nanostructures or influence of the surrounding chemicals might make the absorbance of the sample for the two polarizations deviate from the expectation. This issue would therefore require adjusting the illumination intensity to make sure the temperature increase is the same. The effective absorbance of the sample can be quite easily characterized with an infrared camera, by measuring the temperature increase under the two orthogonal polarizations. Alternatively, one can measure the sample's optical properties with a regular spectrometer, but taking into account that regular transmission measurements typically provide the extinction spectrum and that this can differ markedly from the absorption spectrum because of strong scattering effects in plasmonic nanostructures. The application of this two-polarization procedure to discern photothermal and photochemical effects has not yet been reported.

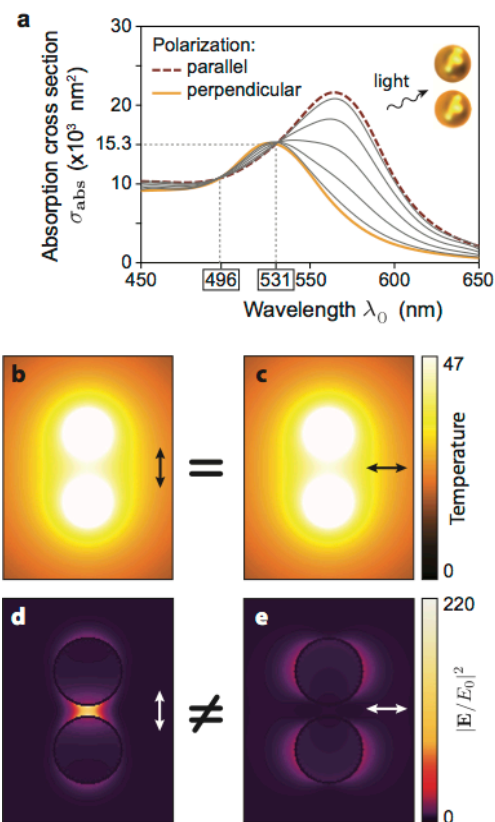

Figure 6: (a) Absorption cross section of a dimer structure composed of two 60 nm gold spheres in water separated by a 10 nm gap, for polarizations of the incident electric field parallel (longitudinal) and perpendicular (transverse) to the dimer axis. Gray lines correspond to intermediate polarization angles (15°, 30°, 45°, 60°, and 75°). A crossing is observed for the isosbestic wavelengths λ = 496 and 531 nm. Temperature maps and corresponding electric field intensity enhancements for (b,d) longitudinal and (c,e) transverse polarizations of the incident light for a dimer illuminated at 531 nm with a light irradiance of 1 mW/µm2. Reproduced with permission from Ref. [80]. Copyright 2018, American Chemical Society.

## Procedure # 7: Comparing the effects of several wavelengths

In the same spirit as Procedure #6, instead of considering two polarizations with the same absorption, one can also consider two wavelengths. If, say, a blue illumination gives rise to a higher chemical rate compared to the red one while the sample temperature increase remains identical (as verified using IR imaging for instance), then optical heating cannot explain the increase of the photochemical rate by itself and it would indicate that the photon energy also matters. A different rate enhancement under different wavelengths would not directly imply a plasmonic hot-carrier driven process, as for example near-field enhancements are not directly proportional to the absorbance of the sample, but it would at least give strong evidence for the existence of a photochemical process. Compared to the two-polarization approach, the drawback of the two-wavelength approach is that it requires two sources of light, ideally with adjustable wavelengths. However, the benefit is that the geometry of the structure is not critical. If the wavelength can be spanned, one can adjust it to find two wavelengths corresponding to approximately identical absorbance.

Sometimes the absorbance of the sample cannot be determined with certainty. In these cases it is possible to record the reaction rate as a function of the optical power $\eta(P)$ under two irradiation wavelengths, one with energetic photons ($\lambda_{\text{blue}}$), which can give rise to hot charge carriers either via interband absorption or plasmon-mediated Landau damping, and one which cannot account for any hot charge carrier effect ($\lambda_{\text{red}}$) and which will necessarily only lead to a thermally-induced process. As the absorbance of the sample has no reason to be the same at $\lambda_{\text{red}}$ and $\lambda_{\text{blue}}$, the line shapes of the "rate versus power" plots at the two wavelengths ($\eta_{\text{blue}}(P_{\text{blue}})$ and $\eta_{\text{red}}(P_{\text{red}})$) will not overlap in principle. However, if the process is purely thermally driven in both cases, then there should exist a constant factor $\alpha$ such that $\eta_{\text{blue}}(\alpha.P_{\text{blue}}) = \eta_{\text{red}}(P_{\text{red}})$ and for which the two plots perfectly overlap. If, on the contrary, photochemical processes are also activated under illumination with $\lambda_{\text{blue}}$, then it should be impossible to find a constant factor $\alpha$ fulfilling this condition.

Another approach that requires the use of a monochromator or of a light source with adjustable wavelength is the acquisition of a spectrum of the chemical rate, as a means to possibly evidence a wavelength threshold for a sharp transition, above which the reaction is markedly damped. This would be a typical feature of a process where the quantum nature of light matters, which is the case for a hot-carrier or a chemical interface damping mechanism, and not the case for a photothermal process, where the chemical rate is supposed to follow the smooth plasmonic resonance absorption spectrum.

## More advanced approaches

So far, we have described relatively simple experimental procedures that should be easy to implement in most laboratories interested in studying photochemical plasmonic effects. Alternative, more sophisticated methods for discriminating between photothermal and photochemical effects in plasmonics exist and have been successfully applied by the community in the past. Without attempting to draw an exhaustive picture of the field, we provide here a brief list of more advanced approaches, namely, temperature microscopies, the kinetic isotope effect and the monitoring of chemical selectivity.

Temperature microscopy techniques have been developed and applied to plasmonics over the last decade [86, 87] and can be used to determine the temperature increase in plasmon-assisted chemistry experiments. Most of the developed techniques are based on optical measurements. Among them, fluorescence measurements are often involved but adding chemicals may not be the best strategy in such experiments. There exist more suitable, label-free techniques such as Raman spectroscopy [88] or fluorescence anti-stokes emission [89] of the metal nanoparticles themselves. The use of NV centers or lanthanide-doped nanoparticle [90, 91] could also give reliable measurements, as these are chemically inert and thermally robust. However, when collective thermal effects are dominant (see also the discussion in the next section), small-scale temperature measurements are not relevant as the temperature distribution is uniform throughout the sample at the macroscale, despite the nanoscale nature of the heat sources.

Another approach is the use of kinetic isotope effects (KIEs), which indicate the change of a chemical reaction rate when replacing an atom of a reactant molecule by one of its isotopes ($^{16}O_2$ by $^{18}O_2$ for instance). KIEs can be used to elucidate reaction mechanisms and have been used in the past for discriminating the effects of temperature and light on the adsorption/oxidation of CO on a ruthenium substrate [92]. In this study, hot-electron effects have been demonstrated, supported by time-resolved measurements enabled by the use of a femtosecond-pulsed laser illumination. The KIE was later used to discriminate the roles of temperature and light in several plasmon-assisted catalysis experiments under cw illumination [51]. Despite the fact that KIE experiments can provide direct evidence for non-thermal effects in photocatalysis, they are typically rather expensive, which may explain why they have not been more widely used in the community.

Finally, there exist chemical reactions that yield different products when assisted by light or by heat. Comparing the products of the reactions following a light excitation and a resistive heating (using a hot plate for instance) can also be a means to show that photothermal effects cannot be the only mechanism at play. This mechanism-dependent reaction selectivity has been recently used for example to discriminate photothermal from photonic effects in methane production, propylene epoxidation, and methane reforming [93, 94, 95].

# Temperature calculation: A risky approach due to collective photothermal effects

An easy misconception in plasmon-driven chemistry that is often encountered in the literature is that wide field illumination of a macroscopic sample can lead to highly localized thermal hotspots at the locations of the metal nanoparticles. Against the common sense, however, if one illuminates a macroscopic distribution of nanoparticles (say over 1 in²) in 2D or 3D samples, it is *not* possible to generate thermal hot spots around each nanoparticle. For instance, for a gold nanoparticle in a water-like medium under illumination, its temperature increase is given by [96]

$$\delta T = \frac{\sigma_{\mathrm{abs}} I}{4\pi \kappa \beta R} \qquad (3)$$

where $\sigma_{abs}$ is the nanoparticle absorption cross section, $I$ the irradiance (power per unit area), $\kappa$ the thermal conductivity of the surroundings, $R$ the effective radius of the nanoparticle (radius of a sphere of identical volume), and $\beta \geq 1$ a correction factor taking into account the shape of the nanoparticle ($\beta = 1$ for a sphere). For a 50-nm diameter nanosphere, to increase its temperature by 1 K, the expression tells that one would need an irradiance $I$ on the order of 0.1 mW/µm². This is possible by focusing a laser, but using wide-field illumination as performed in plasmon-assisted chemistry (for instance 1 inch in diameter), a total light power of 10,000 W would be required. Despite this, equation 3 is very often used in plasmonics to calculate the magnitude of photothermal effects under wide field illumination [9, 34, 69, 28, 93] and naturally yields severe underestimations of the actual sample temperature increase, as it only considers the (negligible) local temperature increase while neglecting the (dominant) collective heating that we shall now explain.

When illuminating an ensemble of nanoparticles, either in a 2D layer [78], or a 3D (liquid or solid) sample [97], the most important parameter is no longer the absorption cross section of the individual nanoparticles, but the absorbance of the sample, i.e., its color (white, dark grey, black, …). If the nanoparticle density is sufficiently high, a temperature increase will be observed. Notably, this temperature increase will be spread across the entire sample and it will be continuous without any nanoscale features. This effect is commonly known as photothermal collective heating or homogenization effect in plasmonics [60, 64].

This result can be counterintuitive from an optics perspective. In most randomly dispersed plasmonic samples, if nanoparticles are separated by a few diameters, they can be considered as optically decoupled, regardless their amount. This reasoning, however, does not apply in thermodynamics, where in addition to the average nearest-neighbor distance $p$ and the particle size $R$, the number of nanoparticles $N$ under illumination strongly matters. The temperature increase experienced by a nanoparticle results from two contributions: its own heat generation, and the heat generated by the other $N - 1$ nanoparticles under illumination in the sample. For a 2D distribution of nanoparticles, like in heterogeneous chemistry where particles are covering a flat substrate, or in photovoltaics, the balance of these two contributions can be estimated using a dimensionless number, $\zeta_{2D} = \delta T_{\mathrm{NP}}/\delta T_{\mathrm{all}}$

[60, 64], indicating the ratio between the local and the collective temperature increase and defined as:

$$\zeta_{2D} \sim p/3R\sqrt{N} = p^2/3RL = (3ARL)^{-1} \qquad (4)$$

where $p$ is the average nanoparticle distance, $R$ the typical nanoparticle radius, $N$ the number of nanoparticles under illumination, $L^2 = p^2 N$ the heated area and $A$ the nanoparticle areal density. This expression assumes uniform and infinite media above and below the layer through which heat escapes.

As an example, from Figure 7 taken from the literature, one can estimate $\zeta_{2D} = p^2/3RL \sim 10^{-4} \ll 1$ (with $R = 7$ nm, $p = 150$ nm, $L = 1$ cm). Such a small $\zeta_{2D}$ value indicates a dominant collective effect, characterized by a uniform sample temperature increase around $10^4$ higher than what can be calculated with the expression (3) of $\delta T$ for an isolated particle. In other words, the temperature increase of a given nanoparticle represented in Figure 7 mostly comes from the heating of the other $N-1$ nanoparticles, although they may seem far away and they are optically decoupled. Figure 7b-d presents numerical simulations related to the practical example given in Figure 7a. Figure 7b plots the heat source density arising from the three particles in the field as under illumination at 2.4 W/cm². Figure 7c plots the calculated temperature distribution as if the three particles were the only ones under illumination. Localized temperature increases can be observed but with extremely small amplitudes. Conversely, Figure 7d displays the temperature distribution considering the illumination of an area of 1 cm², with the same nanoparticle density as in Figure 7a, leading to two striking features: a uniform temperature without hot spots and a much higher temperature increase, around 4 orders of magnitude higher, as predicted by the estimation of $\zeta_{2D}$ above.

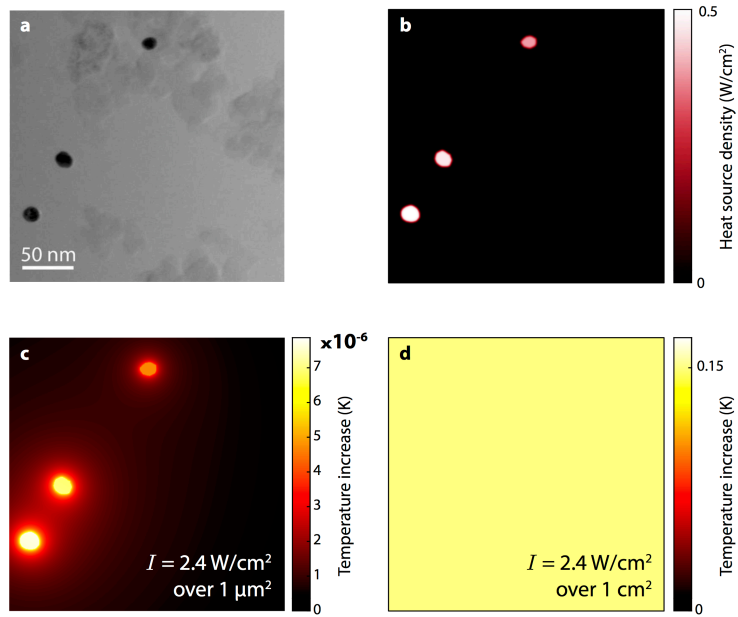

Figure 7: (a) STEM bright field image of 1% Au/SiO$_2$ sample used in Refs. [9, 28] for the plasmon-induced dissociation of H$_2$ on Au. Reproduced from the Suppl. Info. of Ref. [28]. (b) Calculated heat source density associated to (a) assuming an irradiance I = 2.4 W/cm², as in the original work. (c) Temperature distribution calculated using the Laplace Matrix inversion (LMI) method [98], assuming that only these three nanoparticles are illuminated with an irradiance I = 2.4 W/cm². (d) Temperature distribution using Eq. (19) of Ref. [60], assuming a macroscale illumination with an irradiance I = 2.4

W/cm² over a spot area of 1 cm², featuring a perfectly uniform temperature, and around 4 orders of magnitude higher than in (c).

The previous example considered a 2D distribution of nanoparticles. When nanoparticles are distributed in three dimensions, collective thermal effects are even stronger. Let us for instance consider the case of solar light illuminating a test tube (usually through a Fresnel lens to concentrate the light intensity [69, 99]) containing a solution of highly concentrated gold nanoparticles, so dense that it looks dark-grey or even black. In this kind of study, using Eq. (3) can lead to an estimation of the temperature increase as small as 0.04°C, which contradicts the experimental observation of water boiling [69]. Indeed, using Eq. (3) amounts to considering that the system is composed of a test tube containing a transparent liquid in which a single nanoparticle is dissolved, while in reality the system consists of a test tube that contains a very absorbent (black) solution. Illuminating the latter system naturally leads to much higher temperatures. In such studies involving a 3-dimensional system, simulations are more complicated than in 2 dimensions (like in Figure 7). Proper numerical simulations, for instance using the Finite Element Method, should include the full geometry of the system (the absorbing solution and the ice bath [99]) as well as an estimate of the conductive and convective heat and mass transfer in the fluid [35, 100]. For systems in which collective thermal effects lead to a temperature profile that is smooth on the macroscopic scale, using a simple thermocouple (Procedure #3) is the easiest way to accurately and faithfully monitor the temperature of the solution. For systems in which large temperature gradients arise due to inhomogeneous distribution of absorbed optical power, however, multiple thermocouple readings paired with proper modelling of light propagation and heat dissipation can be used to properly account for photothermal plasmonic effects [35, 100].

Finally, let us discuss the case of pulsed laser illumination. Femto- to pico-second illumination can be used as a means to further confine the temperature increase around the plasmonic nanoparticles under illumination [101], but it does not prevent the occurrence of thermal collective effects. Even if the sample is illuminated by fs pulses of light characterized by a fluence $F$ (energy per unit area) and a repetition rate $f$, there still exists an average irradiance $<I> = Ff$ (power per unit area) that contributes to an overall warming of the sample. And the expected temperature increase $\delta T_{\text{NP}}^{\text{pulsed}}$ experienced by a nanoparticle, following a pulse absorption, can be much weaker than the overall temperature increase $\delta T_{\text{all}}$ experienced by the whole sample due to heat accumulation on the macroscale, especially if many nanoparticles are illuminated at once. To quantify one regime or another, there exists a simple dimensionless number quantifying the ratio $\zeta_{\text{2D}}^{\text{pulsed}} = \delta T_{\text{NP}}^{\text{pulsed}} / \delta T_{\text{all}}$ defined as [60]:

$$\zeta_{\text{2D}}^{\text{pulsed}} = \frac{\kappa p^2}{\rho c_p f R^3 L} \quad (5)$$

where $\kappa$ is the thermal conductivity of the surrounding medium (or an average of the different media), $p$ is the typical nanoparticle nearest neighbor distance, $\rho$ and $c_p$ are the mass density and specific heat capacity of the nanoparticle, $R$ the typical size of the particle and $L$ the typical size of the nanoparticle assembly under illumination, usually corresponding to the size of the laser beam.

As an example, in recent reports on heterogeneous catalysis of H₂ dissociation on Al-Pd heterodimers under fs-pulsed illumination [59, 102], the temperature increase $\delta T_{\text{NP}}^{\text{pulsed}}$ was computed using a valid expression, but thermal collective effects were not considered. Based on the experimental details, however, one can calculate $\zeta_{\text{2D}}^{\text{pulsed}} \sim 10^{-4} \ll 1$, meaning that the temperature increase is rather mainly dominated by collective thermal effects.

# Conclusions

The idea to write this article arose from the observation that the plasmonics community is facing some difficulties to properly gauge the contribution of photothermal effects in plasmon-driven chemical reactions, and is currently animated by an active debate. In this context, we here propose simple experimental procedures that can help researchers detect and in some cases quantify photothermal effects in plasmon-assisted chemical reactions. This work is also intended to help readers and reviewers to develop a critical view on this field of research.

Although the present paper mostly focuses on photothermal versus hot-carrier injection processes, other mechanisms of plasmon enhancement of chemical reactions can often be invoked, such as the optical near-field enhancement for photo-activated reactions, often referred to as chemical interface damping [21] or the plasmon induced charge transfer transition (PICTT) [103]. In many experimental settings, several of them could even occur concomitantly. In this context, although they may be less appealing from a fundamental point of view, photothermal effects are not necessarily detrimental as they also contribute to an increase in the reaction rates. In plasmonics-assisted chemistry, however, non-thermal activation mechanisms are often more attractive for at least two reasons: (i) they can increase the selectivity or specificity of a plasmonic catalyst, by activating reaction pathways that are typically thermally inaccessible [94, 104, 93], and (ii) they can increase the activity of a catalyst and accelerate the rate of chemical reactions at milder than usual temperatures, hence preventing undesired effects, such as degradation of catalyst or chemicals (coking) and loss of selectivity. For this reason, in order to ascertain the relevance of using light rather than heat to activate a particular chemical reaction, it is of paramount importance to test whether simple heating of the sample could yield the same results in terms of reaction rate enhancements or selectivity of the products.

An important message in our work is that thermodynamics laws are very different from the physics one usually deals with in plasmonics. For instance, photothermal collective effects in randomly distributed ensembles of nanoparticles, a source of possible misinterpretations as explained in this article, have no equivalent counterpart in optics, making them counterintuitive at first glance. Here, we argue that photothermal processes should be a primary concern of any researcher in plasmonics, as they can lead to misinterpretations in any plasmon-driven process involving the illumination of a large number of nanoparticles for which photothermal effects cannot be ruled out. Although this article was mainly illustrated with examples related to heterogeneous catalysis and photovoltaics, the proposed experimental procedures also apply to gauge photothermal effects in other fields, such as general plasmon-assisted nanochemistry, biosensing, and SERS.

### Acknowledgments

The authors acknowledge fruitful discussion with F. Javier Garcia de Abajo, Ludovic Douillard and Beniamino Sciacca. A. B. acknowledges support by the Dutch Research Council (Nederlandse Organisatie voor Wetenschappelijk Onderzoek) via the NWO Vidi award 680-47-550.

### Conflict of interest

The authors declare that they have no conflict of interest.